\documentclass[twocolumn,english,prl,amsmath,amsfonts,twocolumn]{revtex4}
\usepackage[T1]{fontenc}
\usepackage[utf8]{inputenc}
\usepackage{graphicx}
\usepackage{amssymb}

\makeatletter
\@ifundefined{textcolor}{}
{%
 \definecolor{BLACK}{gray}{0}
 \definecolor{WHITE}{gray}{1}
 \definecolor{RED}{rgb}{1,0,0}
 \definecolor{GREEN}{rgb}{0,1,0}
 \definecolor{BLUE}{rgb}{0,0,1}
 \definecolor{CYAN}{cmyk}{1,0,0,0}
 \definecolor{MAGENTA}{cmyk}{0,1,0,0}
 \definecolor{YELLOW}{cmyk}{0,0,1,0}
 }

\makeatother

\usepackage{babel}

\begin{document}

\title{Exciton-polariton emission from organic semiconductor optical waveguides}

\author{Tal Ellenbogen}

\email{tale@seas.harvard.edu}

\author{Kenneth B. Crozier}

\email{kcrozier@seas.harvard.edu}

\affiliation{Harvard School of Engineering and Applied Sciences, 33 Oxford Street,
Cambridge, Massachusetts 02138, USA}
\begin{abstract}
We photo-excite slab polymer waveguides doped with J-aggregating dye
molecules and measure the leaky emission from strongly coupled waveguide
exciton polariton modes at room temperature. We show that the momentum
of the waveguide exciton polaritons can be controlled by modifying
the thickness of the excitonic waveguide. Non-resonantly pumped excitons
in the slab excitonic waveguide decay into transverse electric and
transverse magnetic strongly coupled exciton waveguide modes with
radial symmetry. These leak to cones of light with radial and azimuthal
polarizations. 
\end{abstract}


\maketitle
The emission of light by recombination of electron-hole pairs (excitons)
in excitonic materials, e.g. semiconductors, quantum wells, dye molecules,
is used for a wide range of applications, including lasers, light
emitting diodes \cite{key-4}, and fluorescent tags in biology \cite{key-6}.
The majority of the applications based on photoemission from excitons
operate in the weak coupling regime, where the exciton is annihilated
and as a consequence a photon is emitted either by spontaneous emission
or by stimulated emission. The exciton and emitted photon can therefore
be treated as two separate physical entities. On the other hand, in
the strong coupling regime, the exciton and photon mode can exchange
energy at a rate faster than the decay rate of the exciton and the
escape rate of the photon, and the two particles create a hybrid state
known as exciton-polariton (EP) \cite{key-7,key-8}. The EP can be
considered as a quasi particle with mixed properties of light and
matter. The masses of cavity EPs are $10^{-4}-10^{-5}$ times those
of typical semiconductor excitons \cite{key-9}. In comparison to
bare photons which cannot scatter each other, EPs can scatter each
other thanks to their excitonic part, leading to strong population-dependent
nonlinearities \cite{key-10,key-11} and interesting physical phenomena,
e.g. Bose-Einstein condensation (BEC) which results in macroscopic
coherence of the condensate and superfluidity \cite{key-12,key-13,key-14}.
Furthermore, these hybrid light matter states have been shown to allow
new device applications such as the generation of low threshold coherent
emission at room temperature without the need for population inversion
\cite{key-10,key-11,key-12,key-13,key-14,key-15,key-16} and low threshold
all-optical switches based on exciton-polariton scattering \cite{key-17}. 

To date, most EP studies and devices have been demonstrated using
optical cavities \cite{key-7,key-8,key-9,key-10,key-11,key-12,key-13,key-14,key-15,key-16,key-17,key-18}.
In optical cavities, strong confinement of the optical modes makes
reaching the strong coupling regime possible, and the dispersion relation
of the cavity EP exhibits a potential well, allowing BEC of the EPs
into the lowest energy state. However, this configuration is somewhat
limited since the EP modes of the system are localized and cannot
propagate, and cannot be probed directly. In addition these devices,
being based on cavities, are usually rigid and are sensitive to deformations
\cite{key-14}. This motivates the investigation of configurations
other than cavities that allow strong coupling of electromagnetic
modes with excitons. 

Recently it was shown that strong coupling can occur between excitons
and surface plasmon polaritons with large room temperature Rabi-splitting
values \cite{key-1,key-19,key-20}. Another optical configuration
that strongly confines photons, and therefore can be considered for
strong coupling with excitons, is an optical waveguide. This configuration
is both common and highly suitable for applications, but has received
only modest attention in the context of strongly coupled modes \cite{key-21,key-22,key-23,key-24}.
Waveguide modes are advantageous compared to cavity modes since they
can propagate, can be studied and directly probed at the near field
through their evanescent tails and the devices can be deformed with
little change to the modes of the system, in the same way that optical
fibers can be bent with little effect. Studying strong coupling in
this configuration could potentially provide new insights into EPs
and could permit new applications and devices. 

We have recently shown that the eigenmodes of a thin polymer waveguide
doped with J-aggregate excitons are strongly coupled waveguide-exciton-polariton
(WGEP) modes \cite{key-25}. This was validated by numerical studies
and experimental reflectivity measurements of the dispersion relation.
The strong coupling phenomenon was seen to manifest itself with large
splitting in the curve of the dispersion relation of both the fundamental
transverse electric (TE) and transverse magnetic (TM) waveguide modes
of the system. We also showed that the strength of the coupling between
the excitons and the TE waveguide modes is similar to that of excitons
and surface plasmon polariton modes in the same system, where the
waveguide mode benefits from lower losses. 

Here we study the emission properties of J-aggregating dye molecules
embedded in polymer optical waveguides. We show that there is strong
leaky emission from WGEP modes to angles corresponding to the phase
matching conditions between the fundamental WGEP modes and leaky photons.
We experimentally confirm that, much like cavity EPs, the coupling
strength between the optical waveguide modes and excitons is proportional
to the square root of the absorbance in the J-aggregates. Therefore,
the large oscillator strengths of the J-aggregate excitons make it
possible to observe this phenomenon at room temperature. We show that,
due to the extended photonic part of WGEP modes, we can control their
momentum by appropriate choice of the thickness of the excitonic polymer
waveguide. We further show that non-resonantly pumped excitons decay
into WGEPs with radial symmetry, which leak to cones of light in free
space that have radial and azimuthal polarizations. 

To fabricate the EP slab waveguide we deposit a thin silver film (40
nm thick) on glass by thermal evaporation, followed by spin coating
an aqueous solution of 5\% polyvinyl alcohol (PVA) mixed in a 3:1
ratio with a 6.6 mM solution of J-aggregating cyanine dye. The cyanine
dye used is the 5,6-dichloro-2-{[}3-{[}5,6-dichloro-1-ethyl-3-(3-sulfopropyl)-2(3H)-benzimidazolidene{]}-1-propenyl{]}-1-ethyl-3-(3-sulfopropyl)
benzimidazolium hydroxide, inner salt, sodium salt, NK2203, Hayashibara
(TDBC), which is known to form linear chains of J-aggregating molecules
at low concentrations (>0.05mM) with large oscillator strengths \cite{key-26}.
The thickness of the spin coated polymer film is measured by a profilometer
to be $\backsim$275 nm (Veeco Dektak 6M). The doped polymer film
acts as the optical waveguide in our system. The thin metal film separates
the polymer waveguide and the glass substrate, enabling optical confinement
in the polymer slab, which would otherwise be difficult due to their
very similar indices of refraction ($\lambda=590$$\: nm$: $n_{BK7}=1.52$,
$n_{PVA}=1.49-1.55$ \cite{key-28}). The metal film is kept thin
enough to allow optical coupling to the waveguide mode from the substrate
side. 

\begin{figure}
\label{Flo:Fig1}\includegraphics{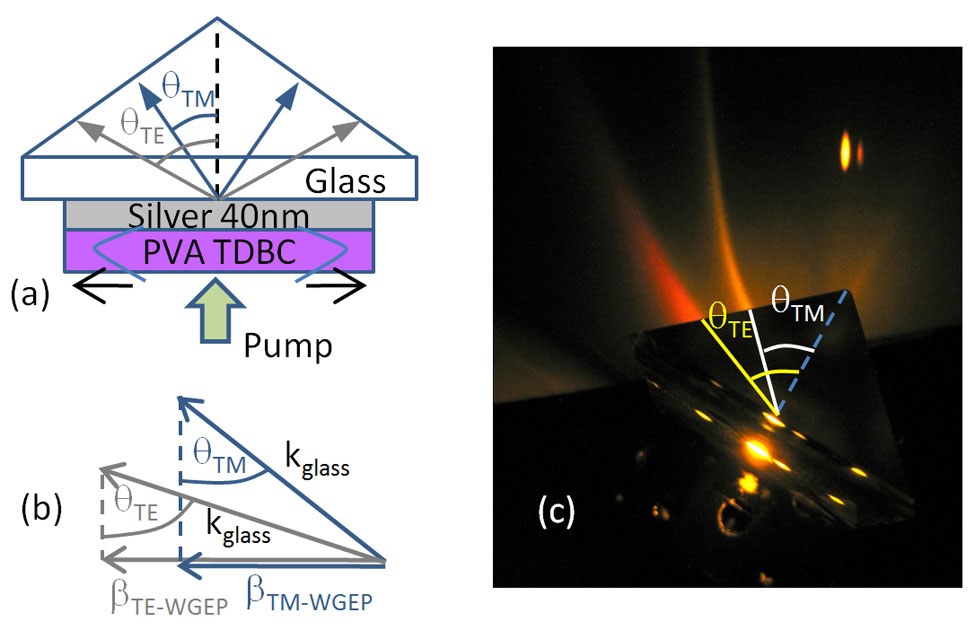}\caption{(color online). (a) Schematic representation of the sample mounted
on a prism. (b) Momentum conservation diagram. (c) Photograph of leaky
emission from the sample projected on a white screen. }

\end{figure}

A schematic representation of the sample mounted on a glass prism
is shown in Fig. 1(a). Index matching oil is used in between the sample
and prism. The excitons are non-resonantly pumped from the waveguide
side and relax into the eigenmodes of the system which are WGEP modes.
These hybrid modes couple out to photons through the prism in accordance
with the momentum conservation condition at the boundary:\begin{equation}
k_{glass}sin(\theta)=\beta_{WGEP}\label{eq:1}\end{equation}
where $k_{glass}=2\pi n_{glass}/\lambda$ is the momentum of the leaky
photons in the glass, $n_{glass}$ is the refractive index, $\lambda$
is the free space wavelength, $\beta_{WGEP}$ is the momentum of the
WGEP and $\theta$ is leakage angle. Fig 1(b) depicts the momentum
conservation conditions for the TE and the TM modes of the system.
TE and TM WGEP modes carry different momenta and thereby couple out
of the sample to different angles through the prism. Fig. 1(c) shows
a photograph of the leaky emission through the prism projected on
a screen when the sample is pumped by a continuous wave (CW) laser
diode with a wavelength of 532 nm. It should be noted that emission
also occurs on the right side of the prism, but cannot be seen in
Fig. 1(c) as the placement of the screen only captures the emission
from the left side. A 550 nm long pass filter is used at the camera
to block scattered light from the excitation source. Emission into
two distinct angles is clearly observed. The two beams are linearly
polarized where the emission into $\theta_{TE}$ is polarized parallel
to the silver film (s polarized) and the emission into $\theta_{TM}$
is polarized in the plane of the illustration shown in Fig. 1(a) (p
polarized). 

To perform far field spectral analysis of the leaky emission from
the sample, the sample is non-resonantly pumped with a 488 nm CW laser
from the polymer film side (see Fig. 1(a)). The leaky emission is
focused by an objective lens into an optical fiber, and input to a
spectrometer with electrically cooled CCD detector (Synapse, Horiba
Scientific). The collection arm is mounted on a rotation stage, enabling
the angle of collection, $\theta$, to be changed. Fig. 2 (a) shows
the leaky emission spectrum collected at $\theta_{TE}=61.2^{\circ}$.
Peaks can be seen at $\lambda$ = 536 nm, 637 nm and 600 nm, corresponding
to leaky emission from the upper polariton mode (UP), lower polariton
mode (LP), and direct emission from the uncoupled J-aggregate excitons,
respectively. Fig. 2 (b) shows emission measured as a function of
angle and wavelength. The emission peaks measured at each can be mapped
to a classical dispersion relation of energy $E$ vs. momentum $k$
where $E=hc/\lambda$ where $h$ is Planck’s constant, $c$ is the
speed of light and $\beta$ is given by \ref{eq:1}. Emission from
the fundamental TE and TM eigenmodes of the system can be identified
in Fig. 2 (b). Both present anti-crossings between the optical mode
and the exciton resonance, at around $\theta=61^{\circ}$ for TE and
around $\theta=45^{\circ}$ for TM. This verifies that the measured
leaky emission is from WGEP modes. We also perform angle resolved
white light reflectivity measurements in a way similar to \cite{key-25}.
We simulate, using Fresnel reflection calculations, the four layer
system comprising the BK7 glass prism, the silver film (40 nm thick),
the PVA film (275 nm thick) doped with excitons, and the semi-infinite
air section adjacent to the PVA film. To model the permittivity of
the J-aggregate doped PVA film, we use a single Lorentzian oscillator
model \cite{key-25} with $\epsilon_{\infty}=2.4$ as the off-resonance
permittivity of the film, $E_{0}=2074\: meV$ as the resonance energy
and $\Delta E=51\: meV$ as the resonance width. 

\begin{figure}
\includegraphics{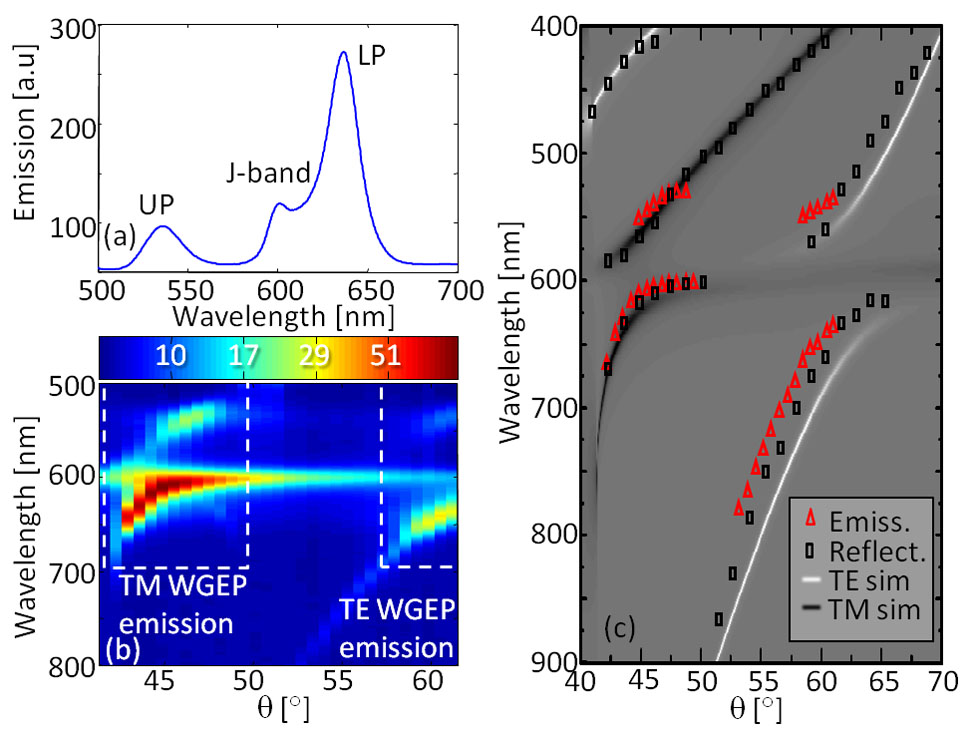}

\caption{(color online). (a) Emission spectrum measured in the far field at
$\theta=61.2^{\circ}$. (b) Angle resolved emission spectra (note
that the color scale is logarithmic with arbitrary units). (c) Splitting
of the dispersion curves of the system. Red triangles show measured
emission peaks, black squares show measured reflectance dips and black
and white curves show simulated angle resolved reflectance dips of
p-polarized and s-polarized light respectively.}

\end{figure}

Fig. 2(c) shows the experimental measurements of emission peaks and
reflectance dips imposed on the simulated reflectance. The red triangles
correspond to emission peaks and the black squares correspond to reflectance
dips. The white and black lines show the simulated reflectance dips
of the TE and TM modes, respectively. The peaks in emission and dips
in reflection measurements agree with the simulated results, verifying
that the leaky emission from the system originates at strongly coupled
WGEP modes.

In order to check the dependence of the coupling strength between
the optical modes and the excitons, $g$, on the absorbance of the
excitonic films, $\alpha$, we fabricate four samples ($\backsim$230
nm thick) with different concentrations of dye molecules at a ratio
of $\backsim$1:2:4:8. White light reflectivity measurements are used
to extract the dispersion relation of the eigenmodes of the different
samples similar to Fig. 2(c) and the coupling strength is measured
on an $E(\beta)$ space by $g=E_{UP}-E_{LP}$ where $E_{UP}$ and
$E_{LP}$ are the resonant energies of the upper and lower WGEP respectively.
The absorbance of the different samples is extracted by performing
optical transmission measurements from the excitonic waveguide deposited
directly on glass. The optical absorbance of the sample is $\alpha=-ln(T/T_{0})$
where $T_{0}$ and $T$ are the transmissions at the J-band through
bare glass and excitonic film on glass, respectively. Fig. 3 plots
the coupling strength as a function of the square root of the normalized
absorbance. A linear relation occurs for the WGEPs, just as it does
for cavity EPs. The inset shows the normalized transmission measurements
from which the absorbance is found. 

Changing the thickness of the exciton doped polymer waveguide presents
a means for modifying the confinement and propagation properties of
the fundamental modes of the waveguide, which in turn affects the
effective indices, $n_{eff}$, of the modes. The momenta of the EPs,
$\beta_{WGEP}=2\pi n_{eff}/\lambda$, can therefore be controlled.
We fabricate samples with excitonic waveguide thicknesses $d$ of
275 nm, 230 nm and 195 nm by choosing the spin coating speed to be
2000 rpm, 2500 rpm and 3000 rpm, respectively. %
\begin{figure}
\includegraphics{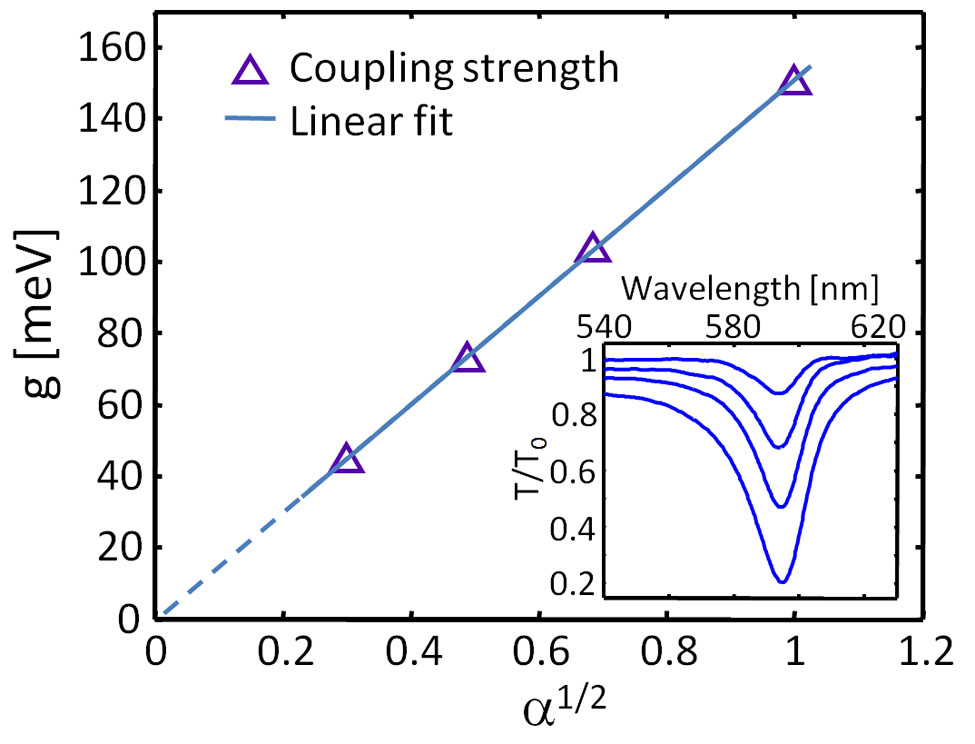}

\caption{(color online). Coupling strength as function of the square root of
the normalized absorbance in the doped PVA film. Inset shows transmittance
measurements from four samples with different dye concentrations corresponding
to the data point shown in the figure. }

\end{figure}
Fig. 4 (a)-(c) present angle resolved leaky emission measurements
from these samples and Fig. 4 (d)-(f) show the corresponding simulated
reflectance results. It is shown that increasing $d$ from 195 nm
to 275 nm results in increasing the leaky emission angle by $\backsim7^{\circ}$,
which is corresponding to $\backsim10\%$ change in the momentum of
the WGEP. 

\begin{figure}
\includegraphics{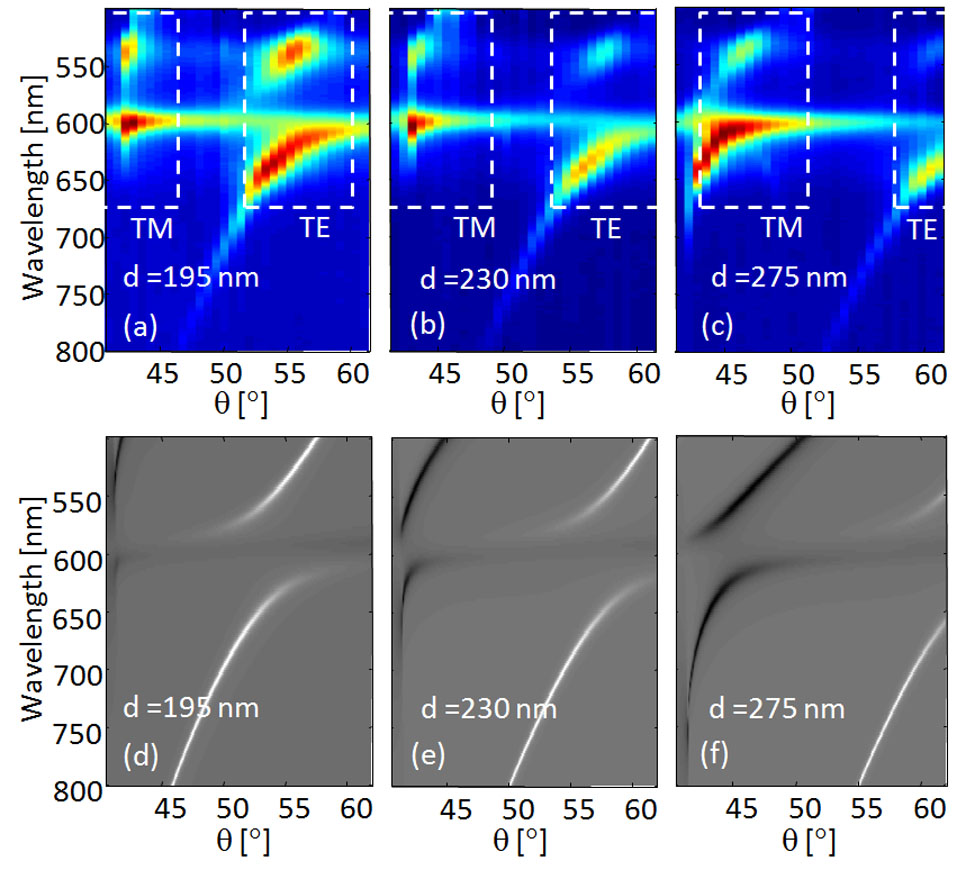}

\caption{(color online). Angle resolved emission spectra from samples with
(a) 195 nm, (b) 230 nm, and (c) 275 nm thicknesses and corresponding
simulated reflectance results (d)-(f) where white lines correspond
to TE modes and black lines to TM modes. }

\end{figure}

Non-resonant excitation of the system from the film side generates
excitons which can relax into WGEPs. The spatial density of WGEP states
in the slab waveguide geometry is radially symmetric. The excitons
therefore relax into radially symmetric WGEPs (with $\beta_{TE}$
and $\beta_{TM}$) which originate at the pump spot. To create an
image of the momentum space of the WGEP modes, we replace the prism
of our set-up with a hemisphere. This enables the WGEPs to couple
to free-space radiation modes, and the angle of emission to be readily
converted to momentum using Eq. 1. Fig. 5(a) depicts the experimental
setup. The sample is pumped with a 532 nm CW laser and the waveguide
modes are coupled out using the glass hemisphere optically coupled
to the sample with index matching oil. The emitted light is filtered
with a 550 nm long pass filter and incident upon a semitransparent
screen. The screen is photographed by a camera from the other side,
with the result shown as Fig 5(b). Two rings corresponding to two
cones of light leaking out of the system can be seen. The inner cone
corresponds to emission from the TM WGEP with $\beta_{TM}$ and the
outer cone corresponds to the emission from the TE WGEP with $\beta_{TE}$.
The cross polarized TM and TE waveguide modes are leaking to cross
polarized light cones where the polarization of the TM WGEP cone is
radial and the polarization of the TE WGEP is azimuthal. Fig. 5(c)
shows the momentum space image obtained when a polarizing film is
placed right after the hemisphere. The white arrow in the picture
shows the direction of the polarization transmitted by the polarizing
film. The crescent, rather than circular, shapes of the WGEP cones
are consistent with the polarizations of the TM and TE WGEPs being
radial and azimuthal, respectively. 

\begin{figure}
\includegraphics{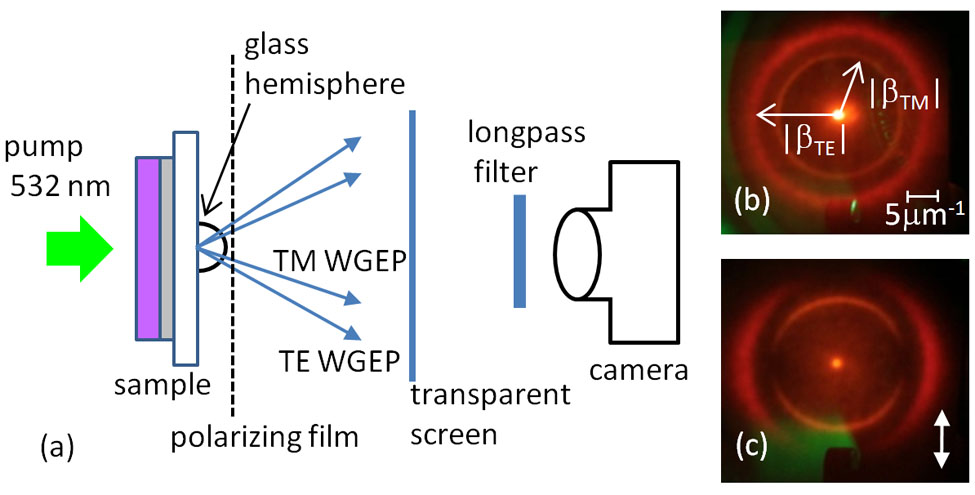}

\caption{(color online). (a) Experimental setup used to measure the momentum
space of the WGEP modes of the system. (b) Photographed momentum space
of WGEPs. (c) Polarization dependence of the momentum space. The white
arrow depicts the polarization of the transmitted light. }

\end{figure}

In conclusion we study leaky emission from thin polymer films doped
with J-aggregate molecules. We show that this emission originates
at strongly coupled WGEP modes, and occurs at angles corresponding
to the phase matching conditions between the strongly coupled light
matter modes and leaky photons. The WGEP modes are studied by emission
and reflectance measurements, which are seen to agree with reflectance
simulations. Similar to cavity EPs, the coupling strength of WGEPs
has a quadratic proportionality to the absorbance in the excitonic
film. We show that the momentum of the waveguide modes can be modified
by changing the thickness of the excitonic waveguide, in a similar
way to regular optical waveguide modes. By capturing the 2D momentum
space of the WGEP in the system, we demonstrate that non-resonant
pumping of the excitons in the system relaxes to WGEPs with radial
symmetry. These modes translate into cones of leaky free space emission
with radial and azimuthal polarization states for the TM WGEP and
TE WGEP respectively. We anticipate that the optical waveguide configuration
with its differences and advantages over the optical cavity configuration
could be used for new fundamental studies on EPs, e.g. by probing
the modes in the near field through their evanescent tails, and will
potentially lead to new applications based on light matter states,
e.g. low-threshold all-optical switching of light in interconnects. 
\begin{acknowledgments}
This work was funded by the Center for Excitonics, an Energy Frontier
Research Center funded by the U.S. Department of Energy, Office of
Science and Office of Basic Energy Sciences under Award Number DE-SC0001088.\end{acknowledgments}


\begin{thebibliography}{24}
\bibitem{key-4} B. E. A. Saleh and M. C. Teich, Fundamentals of Photonics,
2nd ed. (Wiley-Interscience, 2007). 

\bibitem{key-6} J. Zhang et al., Nature Rev. Mol. Cell Biol. 3, 906
(2002).

\bibitem{key-7} C. Weisbuch et al., Phys. Rev. Lett. 69, 3314 (1992). 

\bibitem{key-8} G. Khitrova et al., Nature Phys. 2, 81 (2006). 

\bibitem{key-9} D. M. Whittaker et al., Phys. Rev. Lett. 77, 4792
(1996). 

\bibitem{key-10} F. Tassone and Y. Yamamoto, Phys. Rev. B 59, 10830
(1999). 

\bibitem{key-11} P.G. Savvidis et al., Phys. Rev. Lett. 84, 1547
(2000). 

\bibitem{key-12} H. Deng et al., Science 298, 199 (2002). 

\bibitem{key-13} J. Kasprzak et al., Nature 443, 409 (2006). 

\bibitem{key-14} R. Balili et al., Science 316, 1007 (2007). 

\bibitem{key-15} S. Kena-Cohen and S. R. Forrest, Nature Photon.
4, 371 (2010). 

\bibitem{key-16} S. Christopoulos et al., Phys. Rev. Lett. 98, 126405
(2007). 

\bibitem{key-17} A. Amo et al., Nature Photon. 4, 361 (2010). 

\bibitem{key-18} D. Goldberg et al., Nature Photon. 3, 662 (2009). 

\bibitem{key-1} J. Bellessa et al., Phys. Rev. Lett. 93, 036404 (2004).

\bibitem{key-19} T. K. Hakala et al., Phys. Rev. Lett. 103, 053602
(2009). 

\bibitem{key-20} N. T. Fofang et al., Nano Lett. 8, 3481 (2008). 

\bibitem{key-21}T. Katsuyama and K. Ogawa, J. Appl. Phys. 75, 7607-7625
(1994). 

\bibitem{key-22}L. K. van Vugt et al., Phys. Rev. Lett. 97, 147401
(2006). 

\bibitem{key-23} K. Takazawa et al., Phys. Rev. Lett. 105, 067401
(2010). 

\bibitem{key-24} M. Liscidini et al., Appl. Phys. Lett. 98, 121118
(2011). 

\bibitem{key-25} T. Ellenbogen, P. Steinvurzel and K. B. Crozier,
Appl. Phys. Lett. 98, 261103 (2011). 

\bibitem{key-26} M. van Burgel, D. A. Wiersma and K. Duppen,  J.
Chem. Phys. 102, 20 (1994). 

\bibitem{key-28}D. Cristea et al., Mat. Sci. Eng. C 26, 1049 (2006).
\end{thebibliography}
\end{document}